\documentclass[prb,superscriptaddress,twocolumn,amsmath,amssymb,showpacs]{revtex4}
\usepackage{graphicx}
\begin{document}
\title{Longitudinal and transverse exciton spin relaxation times in single InP/InAsP/InP nanowire quantum dots}

\author{H.\ Sasakura}
    \email[]{hirotaka@eng.hokudai.ac.jp}    
    	\affiliation{Research Institute for Electronic Science, Hokkaido University, Sapporo 001-0021, Japan}

\author{C.\ Hermannst\"{a}dter}

\affiliation{Research Institute for Electronic Science, Hokkaido University, Sapporo 001-0021, Japan}

\author{S.\ N.\ Dorenbos}

\affiliation{Quantum Transport, Kavli Institute of Nanoscience, Delft University of Technology, The Netherlands}

\author{N.\ Akopian}
\affiliation{Quantum Transport, Kavli Institute of Nanoscience, Delft University of Technology, The Netherlands}

\author{M.\ P.\ van\ Kouwen}
\affiliation{Quantum Transport, Kavli Institute of Nanoscience, Delft University of Technology, The Netherlands}

\author{J.\ Motohisa}
\affiliation{Graduate School of Information Science Technology, Hokkaido University, Sapporo 060-0814, Japan}

\author{Y.\ Kobayashi}
\affiliation{Graduate School of Information Science Technology, Hokkaido University, Sapporo 060-0814, Japan}
\author{H.\ Kumano}
       	\affiliation{Research Institute for Electronic Science, Hokkaido University, Sapporo 001-0021, Japan}
\author{K.\ Kondo}
	\affiliation{Research Institute for Electronic Science, Hokkaido University, Sapporo 001-0021, Japan}
	
\author{K.\ Tomioka}
\affiliation{Graduate School of Information Science Technology, Hokkaido University, Sapporo 060-0814, Japan}
\affiliation{Research Center for Integrated Quantum Electronics (RCIQE), Hokkaido University, Sapporo 060-8628, Japan}

\author{T.\ Fukui}
\affiliation{Graduate School of Information Science Technology, Hokkaido University, Sapporo 060-0814, Japan}
\affiliation{Research Center for Integrated Quantum Electronics (RCIQE), Hokkaido University, Sapporo 060-8628, Japan}

\author{I.\ Suemune}
\affiliation{Research Institute for Electronic Science, Hokkaido University, Sapporo 001-0021, Japan}

\author{V.\ Zwiller}
\affiliation{Quantum Transport, Kavli Institute of Nanoscience, Delft University of Technology, The Netherlands}

\date{\today}

\begin{abstract}

We have investigated the optical properties of a single InAsP quantum dot embedded in a standing InP nanowire. A regular array of nanowires was fabricated by epitaxial growth and electron-beam patterning. The elongation of transverse exciton spin relaxation time of the exciton state with decreasing excitation power was observed by first-order photon correlation measurements. This behavior is well explained by the motional narrowing mechanism induced by Gaussian fluctuations of environmental charges in the InP nanowire. The longitudinal exciton spin relaxation time was evaluated by the degree of the random polarization of emission originating from exciton state confined in a single nanowire quantum dots by using Mueller Calculus based on Stokes parameters representation. The decreasing of random polarization component with decreasing excitation power was caused by suppression of exchange interaction of electron and hole due to optically induced internal electric field by the dipoles at the wurtzite and zinc-blende heterointerfaces in InP nanowire. 

\end{abstract}
\maketitle

\section {Introduction}

Semiconductor quantum dots (ODs) are called artificial atoms because of their atom-like discrete electronic structure due to quantum confinement. Exciton spin confined in their structure can relax via a wide variety of interaction and/or scattering: valence band mixing~\cite{Leger07,Belhadj10,Ohno11}, Coulomb interaction with environmental charge fluctuation~\cite{Ber06,Fav07,Pat08}, nuclear spin fluctuation~\cite{Merkulov02,Xu09}, and so on~\cite{Superlattices,Maialle93}. In narrow bandgap semiconductor nanostructures, heavy hole-light hole mixing is enhanced by strong spin-orbit interaction. Recently, research on semiconductor nanowires (NWs)~\cite{Lie02,Xia01} has been revived because of their great potentials for nano-scale electrical and optical devices such as quantum logic circuits, photon detectors~\cite{Wan01}, and photon sources~\cite{Gud05} for quantum information processing~\cite{Aws02} due to the controllability of doping, shape, position, composition, and the possibility of electrical nano-contacts with the improvements in growth and nano-fabrication technology. Fast dephasing of carriers in NW structures fabricated by top-down techniques such as photolithography and electron-beam lithography is improved by the reduction of surface traps induced by defects and/or disorders with epitaxially grown bottom-up single or multi core(-shelled) NW structures~\cite{Mohan05}. QDs embedded NWs (hereafter called NW-QD), i.e. combined 0-D. and 1-D. structures, are among of the most promising candidates for realizing single-photon detector~\cite{Kouwen10} and on-demand single-photon sources~\cite{Reimer11} with high (long) coherence (dephasing). Emitter dephasing and polarization instability set up bit errors and debase communication distance limit when implementing phase and polarization coding quantum key distribution protocol~\cite{Aws02}. In this paper we investigate the transverse and longitudinal exciton spin relaxation times. The transverse exciton spin relaxation time of our NW-QD caused by fluctuation in surrounding excess charges in the NW is suppressed with decreasing excitation power, which was observed by first-order photon correlation measurements under non-resonant excitation condition. The longitudinal exciton spin relaxation time was evaluated by the degree of the random polarization component in time-integrated photoluminescence spectra. The degree of random component is obtained by Mueller analysis of experimental results of full polarization measurement based on Stokes parameters. A decreasing random polarization components of emitted photon from exciton state with decreasing emission energy was observed due to the suppression of overlap integral of electron and hole envelope wave function.

\section {EXPERIMENTS}
\begin{figure}
\begin {center}
\includegraphics[width=\columnwidth]{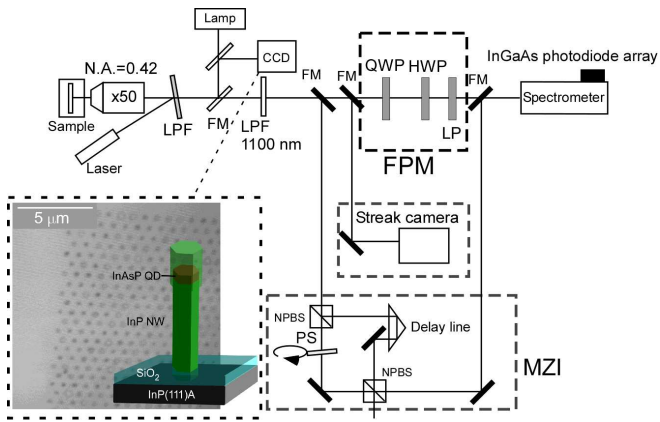}
\caption{ (Color online) Schematic of the experimental setup: First-order photon correlation measurement was performed by measuring the PL passing through a Mach-Zehnder interferometer (MZI). Photon count is displayed as a function of the optical path delay. Full polarization measurement (FPM) was performed by using array of QWP, HWP and LP. Decay time constant of PL was measured by streak camera. LPF:long wave pass filter. FM:flipper-mounted mirror. PS:phase shifter. NPBS:nonpolarized beam splitter. Inset : Microscope image of the sample surface and schematic of the NW-QD grown on InP(111)A substrate. 
}
\label{fig0}
\end {center}
\end{figure}

Arrays of InAsP QDs embedded in InP NWs were synthesized by selective area metalorganic vapor phase epitaxy (SA-MOVPE, Ref.~\onlinecite{Mohan05}). A (111)A InP wafer was covered by 30 nm of SiO$_2$. By electron beam lithography and wet-etching, 40 to 60 nm diameter openings were created to form nanowire nucleation-sites (Fig.~\ref{fig0}). At a growth rate of 3 nm/s, a first 1 $\mu$m long segment of InP was grown by adding trimethylindium (TMI) and tertiarybutylphosphine (TBP) to the MOVPE reactor at 640 $^\circ$C. To form the QDs the temperature was lowered to 580 $^\circ$C and arsine (AsH$_{3}$) was added to the reactor (V/III ratio 340, partial pressure TBP:AsH$_3$ 3:1). At a growth rate of 3 nm/s a 8 to 10 nm layer of InAsP was created forming the QD. To cover the QDs with an InP shell, InP growth was first performed at 580 $^\circ$C. To finalize the NW-QDs, a second 1 $\mu$m segment of InP was grown at 640 $^\circ$C (Fig.~\ref{fig0}). 
Single NW-QD photoluminescence (PL) spectroscopy was performed by standard micro-PL measurements in the far field using a microscope objective as shown in Fig.~\ref{fig0}. The sample was cooled to 5 K in a $^4$He flow cryostat. 
Excitation laser beam travelling along the NW growth direction was focused on the sample surface by a microscope objective. The NW-QD emission collected by the same objective lens was dispersed by a double (triple) grating spectrometer ($f=1.0$ m) and was detected with a liquid-nitrogen-cooled InGaAs photodiode array. The typical exposure time was 1 s to obtain a spectrum with a high signal-to-noise ratio.

\section {Basic optical properties}
\begin{figure}
\begin {center}
\includegraphics[width=\columnwidth]{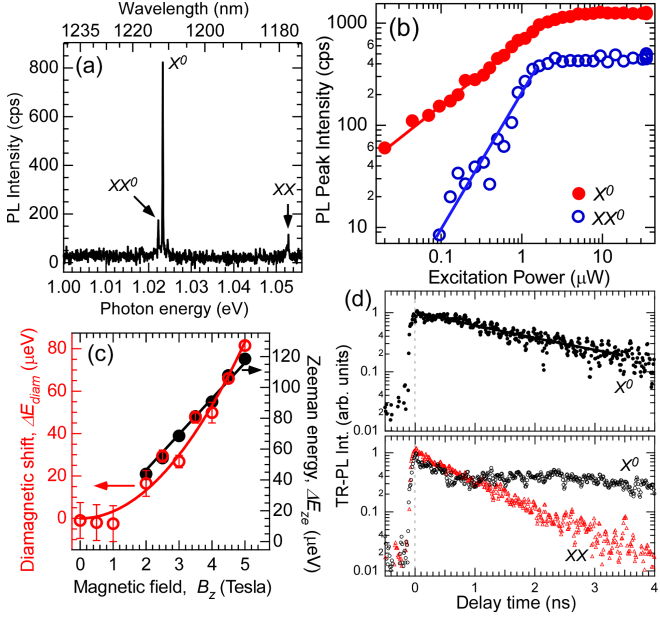}
\caption{(Color online) (a) PL spectrum of a single QD under pulsed TIS laser excitation with excitation power of 1 $\mu$W. (b) PL intensities of $X^{0}$ and $XX^0$ peaks as a function of excitation power. Red and blue lines are guides to the eye. (c) Zeeman energy and diamagnetic shift are shown. Closed (black) circles and open (red) circles represent experimental data, and the (black) line and (red) curve are fitted with $|g_{\textrm {ex}}|\mu _{\textrm B}B_{z}$ and $\gamma _{2}B^{2}_{z}$, respectively. $|g_{\textrm {ex}}|=0.4$ and $\gamma_{2}=23.2\pm 0.3$ $\mu$eV/T$^{2}$. (d) Time-resolved PL of $X^{0}$ (closed (black) circles) and $XX$ (open (red) triangle). Excitation power is set at 0.1 $\mu$W(Upper panel) and 2 $\mu$W(lower panel). The $X^0$ decay time of $\sim$2 ns is fitted with a single exponential function (solid black) line.}
\label{fig1}
\end {center}
\end{figure}

Figure~\ref{fig1}(a) shows a time-integrated PL spectrum measured on a single NW-QD. The excitation has been carried out with a pulsed-Ti:sapphire laser at a wavelength of $\sim$800 nm with a repetition rate of 76 MHz. At low excitation intensity, the $X^0$ is centered at $\sim$1.023282 eV with 46 $\mu$eV FWHM. The fine structure splitting normally induced by QD anisotropy and associated exchange interaction~\cite{Bayer02, Aki05} could not be observed in our measurement system. The spectral resolution that determines peak energy was observed to be less than 5 $\mu$eV by using spectral fitting. Absence of exciton splitting is expected for InAs QDs in InP NWs grown in the (111) direction~\cite{Sin09} and would be ideal for the generation of entangled photons. With increasing excitation power, the PL intensity of $X^0$ linearly increases and saturates at an excitation power of $\sim$2 $\mu$W and an additional line labeled as $XX^0$ appears with its PL intensity increasing super-linearly as shown in Fig.~\ref{fig1}(b). In Fig.~\ref{fig1}(c), Zeeman energy $(\Delta E_{ze})$ and diamagnetic shift $(\Delta E_{diam})$ of a single NW-QD are plotted against the external magnetic field in the growth direction (Faraday geometry). By fitting with $g_{ex}\mu _{\textrm B}B_{z}$ and $\gamma _{2}B^{2}_{z}$,~\cite{Walck98}, we obtained an exciton $g$ factor and diamagnetic coefficient as $|g_{ex}|=0.4$ and $\gamma _{2}=23.2\pm0.3$ $\mu$eV/T$^{2}$, respectively. $\mu_{\textrm B}$ is Bohr magneton. $\gamma _{2}$ is good indicator for the degree of lateral confinement potential. There are some reports of small $\gamma _{2}$ value $\ll 10$ $\mu$eV in type-I InAs/GaAs QD~\cite{Nakaoka04,Schulhauser02} and InP/GaAs QD~\cite{Hayne00,Sugisaki02}. The observed $\gamma _{2}$ is larger than type-I QD and smaller than type-II InP/GaAs QD~\cite{Bansal09}. Bohr radius is deduced at $\sim$7 nm using effective mass of InP. Figure ~\ref{fig1}(d) shows time-resolved PL measurements using a streak camera (Hamamatsu: C9510-NIR). The $X^{0}$ decays exponentially and evaluated decay time constant is $\tau _{r}\sim$2 ns by the fitting (solid line in Fig.~\ref{fig1}(d)), implying that observed NW-QD emission $X^{0}$ is type-I. Furthermore there is an excited state transition $(XX)$ $\sim$30 meV above the $X^0$, which is estimated by the state filling effect in excitation power dependence of time-resolved PL measurements (lower panel of Fig.~\ref{fig1}(d)) and time-integrated PL (not shown). These observed results and quantum disk like structure (height:$\sim$3 nm, diameter:$\sim$100 nm)~\cite{Dorenbos10} indicate that the $X^{0}$ is localized in type-I confinement potential and relatively-weak lateral confinement of InAsP QD embedded in InP NW. In what follows we focus on the $X^{0}$, which generates the single photon state estimated by the second-order photon correlation measurements~\cite{Dorenbos10}.

\section {Excitation power dependence of dephasing}

The Mach-Zehnder interferometer inserted in the optical path shown in Fig.~\ref{fig0} was used to perform first-order photon correlation measurements on the single NW-QD $X^0$ emission. This is a type of time-domain spectroscopy called single-photon Fourier spectroscopy first demonstrated by Kammerer {\it et al}~\cite{Kam02}. Fourier spectroscopy is an interesting method to explore the spectral dynamics of a single transition with both high temporal and spectral resolutions and very low photon losses~\cite{Kam02,Zwi04,Kur06,San02,Ada07,Kam08}. PL passing through the interferometer was dispersed by the double grating spectrometer equipped with an InGaAs photodiode array. The continuous-wave photo-excitation was carried out by a He-Ne laser for the NW barrier excitation. Rotating a thin glass plate (PS in Fig.~\ref{fig0}) set in one of the interferometer arms gave fine tuning of the relative phase between the two arms.

\begin{figure}
\begin{center}
\includegraphics[width=\columnwidth]{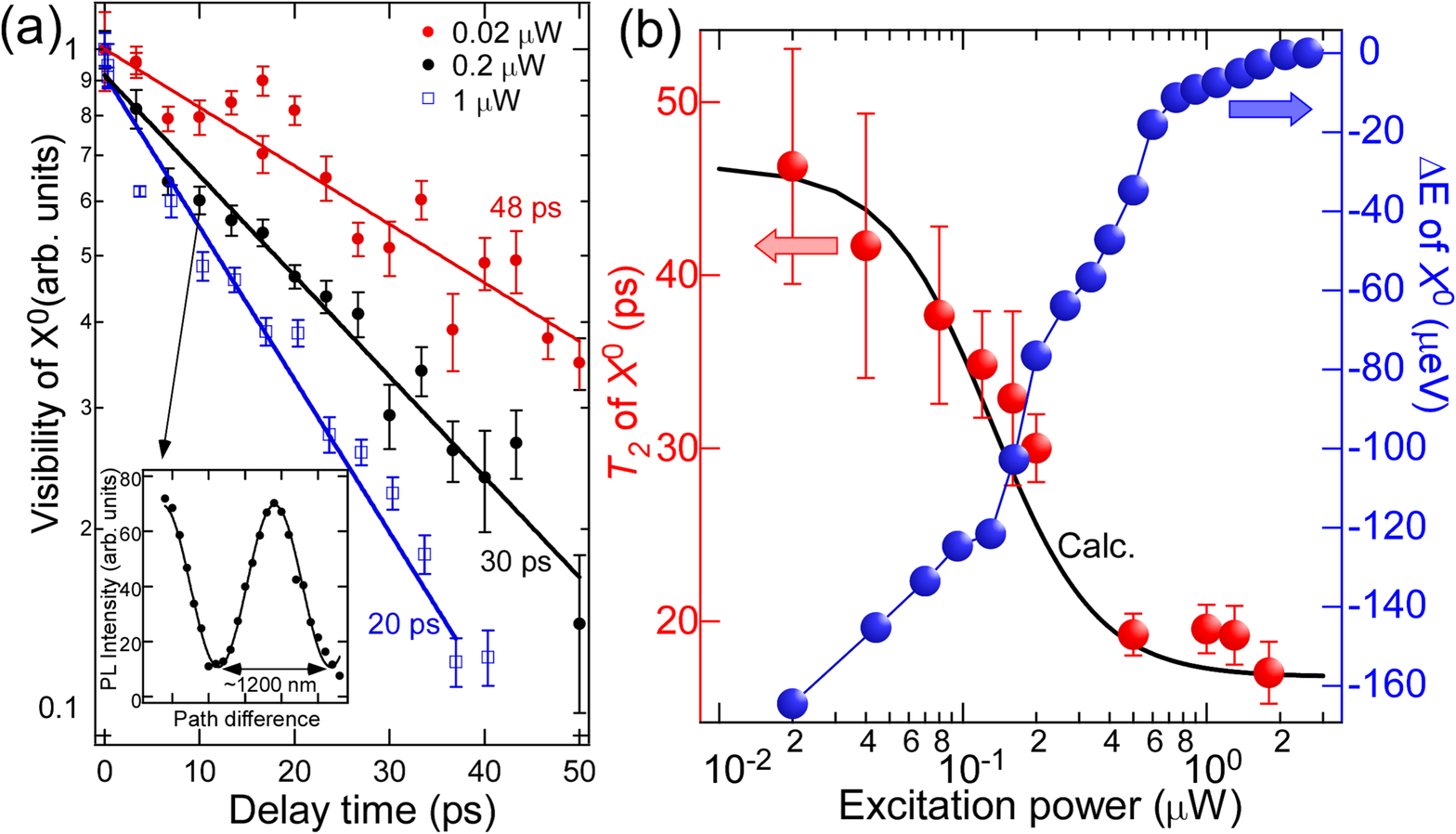}
\caption{(Color online)(a) Visibility plot for a single-dot PL. Inset: Short-period fringe evolution of a single-dot PL at delay time of 10 ns. (b) Excitation power ($P$) dependence of $T_{2}$ (solid (red) circles) with He-Ne laser excitation. Solid (black) line is calculation results using the motional narrowing fluctuation. PL peak energy of $X^0$, $\Delta E =E(P)-E(2\ \mu {\rm W})$, (solid (blue) circles) are plotted. }
\label{fig3}
\end{center} 
\end{figure}

Short-period fringe evolution of the single NW-QD $X^0$ exciton PL emission is shown in the inset of Fig.~\ref{fig3}(a) at around $\tau =10$ ps. The fringe evolution as a function of the delay $\tau$ is given by $I(\tau)=I_{0}[ 1+V(\tau)\cos(E_{0}\tau /\hbar +\theta)]$, where $E_{0}$, $I_{0}$, $V(\tau)$, and $\theta$ are the central detection energy ($\sim$1.023282 eV for $X^0$), the averaged signal intensity, the visibility contrast given by the modulus of the Fourier transform of the intensity spectrum and the phase, respectively. By varying the time delay $\tau$, interference fringes of the single-photon events are recorded. The visibility contrast $V(\tau )$ decays with increasing delay time $\tau$ between the two arms, and the coherence time is measured from the decay contrast of $V(\tau)$. 
The visibility plots shown in Fig.~\ref{fig3}(a) decay following an almost simple exponential function in the investigated range of the excitation power, suggesting that the spectral diffusion phenomenon is in the unconventional motional narrowing regime~\cite{Ber06,Fav07,Pat08}, $\Sigma \tau _{\rm f}/\hbar < 1$, where $\Sigma =2\Delta E_{s}\tau_{\textrm f}/\sqrt{\tau _{\textrm {esc}}\tau _{\textrm {cap}}}$ is the modulation amplitude caused by the individual point charge. $\Delta E _{s}$ is the saturation value~\cite{Fav07}. The rate of correlation between the confined exciton and point charges in the NW is expressed by the combination of capture and escape rates, $1/\tau _{\rm f}=1/\tau _{\rm cap}+1/\tau _{\rm esc}$. Fluctuations of the exciton are induced by the capture (escape) of residual charges by (from) impurities or defects in the NW. For simplicity, we adopt the approach of Favero \textit{et al.}, the acoustic (optical) phonon assisted capture and escape rate is expressed as~\cite{Fav07},
\begin {eqnarray}
\frac {1}{\tau _{\textrm {cap}}}&=&\frac {1}{\tau _{1}}( 1+n_{1}(T) )+\frac {1}{\tau _{2}}(1+n_{2}(T)), \\
\frac {1}{\tau _{\textrm {esc}}}&=&\frac {1}{\tau _{1}}n_{1}(T)+\frac {1}{\tau _{2}}n_{2}(T)+\frac {1}{\tau _{3}}\frac {P^{2}}{P^{2}+P_{0}^{2}}, 
\end {eqnarray}
where $n_{1(2)}$, $P$, and $P_0$ are the Bose-Einstein occupation factors of acoustic (optical) phonon given by $1/(\exp (E_{1(2)}/k_{\textrm B}T)-1)$, the excitation power, and the saturation excitation power of the Auger processes appears, respectively. $E_{1}$ and $E_{2}$ are an acoustic phonon mean energy and optical phonon energy. In this regime the observed transverse exciton spin relaxation time $T_{2}$ is given by $1/T_{2}=(\Sigma /\hbar )^{2}\tau _{\rm f}$ and the line shape remains Lorentzian under Gaussian fluctuations. With decreasing excitation power, the coherence time increases and saturates at $\sim$50 ps as shown in Fig.~\ref{fig3}(b). This suggests that the environmental charge fluctuations were sufficiently suppressed under weak excitation and the observed $T_{2}$ is limited by the relaxation processes in the QD associated with acoustic and/or optical phonons. In Fig.~\ref{fig3}(b), we observe agreements between the observed $T_{2}$ and the calculation (solid (black) line) using similar parameters: $\Delta E_{s}=180$ $\mu$eV, $\tau _{1}=300$ ps, $\tau _{2}=15$ ps, $\tau _{3}=1500$ ps, $P_{0}=220$ nW, $E_{1}=1$ meV, $E_{2}=42.4$ meV as Ref.~\onlinecite{Ber06,Fav07,Pat08}. The prolonged transverse exciton spin relaxation time and deduced value of $\Delta E_{s}$ correspond to the red-shift, $\Delta E$ of $X^{0}$, and the saturation of $\Delta E$ as shown in the right axis of Fig.~\ref{fig3}(b), respectively. Although there are some reports that decay time constant of InP NW with wurtzite crystal structure is order of 100 ns~\cite {Mishra07,Masumoto11}, decay time constant of InP NW part of our sample is $\sim$13 ns~\cite{Sasakura09} (not shown). Therefore this energy shift is attributed to the Stark effect due to the internal electric field induced by the dipoles at the wurtzite and zinc-blende heterointerfaces in the NW~\cite{Pal08} and to screening of the internal field with the photo-excited carriers in the InP NW barrier.

\begin{figure}
\begin {center}
\includegraphics[width=\columnwidth]{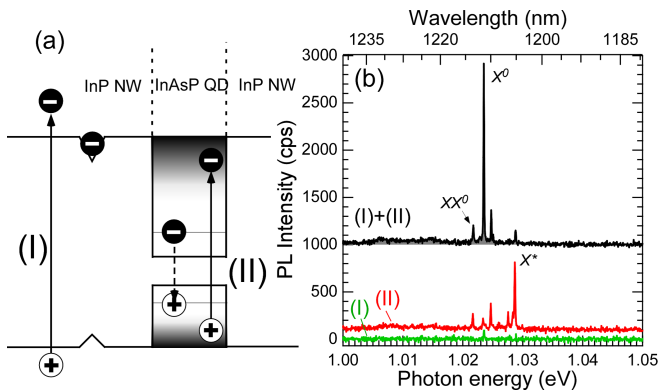}
\caption{ (Color online) (a) Schematic energy band diagram. (I): A green laser (2.33 eV) creates electron and hole in InP NW. (II): cw-TIS laser (1.35 eV) creates exciton in a InAsP NW-QD. Graded areas in InAsP QD indicate continuum state. (b) PL spectra for a InAsP NW-QD obtained with different excitation condition. The laser powers are 10 nW (I) and 25 $\mu$W (II). The emission energies of $X^0$ and $XX^0$ correspond to those in Fig.~\ref{fig1}(a).}
\label{figtwocolor}
\end {center}
\end{figure}

In order to estimate this screening effect of the internal field in InP NW region, we compared PL spectra with two-color excitation method~\cite{Gotoh11} as shown in Figure~\ref{figtwocolor}. We used a cw-diode laser (wavelength: 532 nm), which is above the bandgap energy of InP NW, and a cw-TIS laser (wavelength : 920 nm), which is below the bandgap energy of InP NW, for the two-color excitation. The excitation energy of cw-TIS laser is still above the discrete levels in InAsP NW-QD and creates electron-hole pairs in continuum state~\cite{Toda98,Toda99,Finley01,Kammerer01,Vasanelli02}, which is estimated by the photoluminescence excitation measurements (not shown). Although the $X^{*}$ PL is dominant in case of excitation condition (II), the $X^{0}$ PL appears under the two-color exciton (II)+(I), which corresponds to the PL spectra in Fig.~\ref{fig1}(a). Note that the excitation power of green laser (I) is low enough not to observe PL emission, suggesting that created carriers by green laser are almost used for the screening of the internal field~\cite{Add3}. Therefore $\Delta E$ is attributed to good indicator of the degree of overlap integral of electron and hole envelope wave function.

\section {Longitudinal spin relaxation time}
\begin{figure}
\begin {center}
\includegraphics[width=\columnwidth]{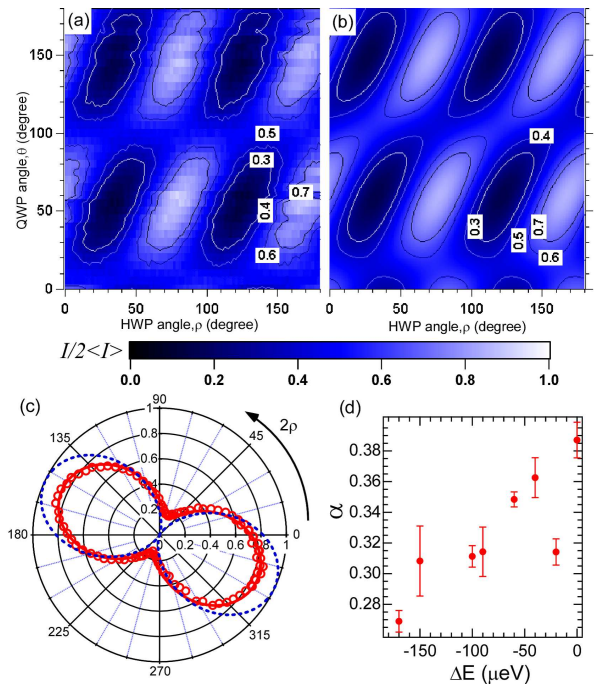}
\caption{(a)2D plot of normalized PL intensities of $X^0$ as a function of $\rho$ and $\theta$. (b) Mueller calculus's results by using $\{ 1, 0.32, -0.6, 0.14 \}$. (c) Polar plot of the PL intensity of $X^0$. The detection angle ($2\rho$) dependence of $I_{\textrm {PL}}(59^{\circ},\rho)/2{\cal h} I_{\textrm {PL}}^{59^{\circ}}(\rho) {\cal i}$ (open (red) circles). Solid (red) and dashed (blue) lines are Mueller calculus's results at $\alpha=0.3$ and $\alpha=0$ by using $\{1, 0.32/0.7, -0.6/0.7, 0.14/0.7 \}$, that is, perfectly coherent light with fixed ratio in three polarization components. All angles are measured in the same laboratory frame. (d) $\Delta E$-dependence of $\alpha$ under fixed Stokes parameters $(S_{1}, S_{2}, S_{3})$ of polarization components. }
\label{figstokes}
\end {center}
\end{figure}

In order to investigate the longitudinal exciton spin relaxation time of the single NW-QD, a quarter wave plate (QWP), half wave plate (HWP), and linear polarizer (LP) were placed in front of the spectrometer as shown in Fig~\ref{fig0}. The transmission axis of the LP is set horizontal in the laboratory frame. $\theta (\rho )$ is the relative angle of Q(H)WP's fast axis to transmission axis of the LP. A He-Ne laser was used as a quasi-randomly polarized excitation source and was focused on the sample surface by the microscope objective. Figure~\ref{figstokes}(a) shows a two-dimensional (2D) plot of normalized PL intensities ($I_{\textrm {PL}}(\theta,\rho)/2{\cal h} I_{\textrm {PL}}^{\theta}(\rho) {\cal i}$) as a function of $\rho$ and $\theta$. This checkered flag-like pattern signifies that observed PL is elliptically-polarized. There are theoretical and experimental reports that the elliptical polarization of photon originating from exciton of a single QD was well explained by the the anisotropic heavy hole-light hole mixing induced by the inhomogeneous piezoelectric field distribution~\cite{Poem07,Leger07,Belhadj10,Ohno11}. Although the amplitude of vibration reaches the maximum at $\theta \approx 59^{\circ}$, which is the cancellation condition of circular polarization components by QWP, the observed amplitude is smaller than 1 (Figure~\ref{figstokes}(c)), suggesting that observed photon state includes random polarization components.

To analyze the obtained results of $X^0$, we introduce normalized Stokes parameters~\cite{Shurcliff,Hecht} with random polarization component, $\{ 1, S_{1}(\alpha), S_{2}(\alpha), S_{3}(\alpha) \}$, where $S_{i}(\alpha)=(1-\alpha)S_{i}$ and $\alpha$ is the degree of random polarization (DRP): $\alpha \{ 1,0,0,0\}$, and $S_{i}$ is Stokes parameters of the coherent components. Applying the traditional four-by-four Mueller matrices, the polarized light propagating can be simulated. Here we assume that three polarization optics are ideal common in the Mueller calculus. The intensity passing through three polarization optics is expressed as,
\begin {eqnarray}
I&=&\frac {1}{2} \left [1+(1-\alpha)\sqrt{S_{3}^{2}+b(\theta)^{2}}\cos(4\rho-2\delta(\theta))\right ],
\label{eq2}
\end {eqnarray}
where $b(\theta)=S_{1}\cos(2\theta)+S_{2}\sin(2\theta)$ and $\delta(\theta)(=0.5\arctan(S_{2}^{'}/S_{1}^{'}))$ is tilting angle from the horizontal axis of linear polarization components after the light passed through QWP($\theta$): ${\textrm {QWP}}(\theta)\{1,S_{1},S_{2},S_{3} \} \rightarrow \{ 1,S_{1}^{'},S_{2}^{'},S_{3}^{'} \}$.

At the cancellation condition of circular polarization components by QWP: $\theta=0.5\arctan (S_{2}/S_{1})$, the amplitude of vibration reaches the maximum. Substituting $\theta=0.5\arctan (S_{2}/S_{1})$ and $\alpha=0$ into Eq. (\ref{eq2}), the amplitude is exactly 1. Note that this is suitable only as a limiting case that the exciton spin-flip time, in other words the longitudinal exciton spin relaxation time $T_{1}$ is much longer than the recombination time $\tau _{r}$. Figure~\ref{figstokes}(c) shows the polar plot at $\theta=59^{\circ}$ of the observed $I_{\textrm {PL}}(59^{\circ},\rho)/2{\cal h} I_{\textrm {PL}}^{59^{\circ}}(\rho) {\cal i}$. The maximum amplitude, $0.85=1-\alpha/2$ is smaller than 1. By using $\alpha =0.3$ and $0.5\arctan (S_{2}/S_{1})=59^{\circ}$, we can deduce the Stokes parameters. Assuming the Stokes parameters $\{ 1, 0.32, -0.6, 0.14 \}$, we can reproduce not only amplitude of $I_{\textrm {PL}}(\theta,\rho)/2{\cal h} I_{\textrm {PL}}^{\theta}(\rho) {\cal i}$ but also its pattern very well as shown in Figure~\ref{figstokes}(b) and (c). 
The observed DRP is evaluated at $0.3\left ( =1-\sqrt {S_{1}(\alpha)^{2}+S_{2}(\alpha)^{2}+S_{3}(\alpha)^{2}} \right )$, suggesting that $T_{1}$ is $\sim$0.7 times shorter than the $\tau _{r}$ of $X^0$ (lower panel of Fig.~\ref{fig1}(d)). Note that the random polarization component can be observed in the time integrated PL as a case of $T_{1}\le \tau _{r}$. The ratio between $T_{1}$ and $\tau _{r}$ is expressed as $T_{1}/\tau _{r}=1-\alpha$. 
In general, the spin of exciton in quantum well at low temperature can mainly flip via the short- and long-range exchange interactions, that is, the Maialle-Andrada-Sham mechanism~\cite{Maialle93}. Figure~\ref{figstokes}(d) shows $\Delta E$-dependence of $\alpha$ under assumption of fixed ratio in three polarization components. $\alpha$ decays slightly with decreasing $\Delta E$, implying that the overlap integral of electron and hole envelope wave function, ${\cal h}\varphi _{e}{\cal j}\varphi _{h}{\cal i}$, decreases due to the internal field~\cite{Pal08}. Because the ratio of $T_{1}/\tau _{r}$ depends on the overlap integral of electron and hole envelope wave function, $T_{1}/\tau _{r}\propto {\cal h}\varphi _{e}{\cal j}\varphi _{h}{\cal i}^{2-\nu}$. Both of observed results of $\nu \ge 2$ and high degree of linear polarization $0.98(=\sqrt{S_{1}^{2}+S_{2}^{2}})$ suggest that the observed $T_{1}$ is dominant with short-range exchange interaction corresponding to strong light hole-heavy hole mixing.

\section {SUMMARY}
In conclusion, we investigated optical properties of a single quantum dot embedded in a vertically standing nanowire structure and measured the transverse and longitudinal exciton spin relaxation times. The transverse exciton spin relaxation induced by the fluctuations in the environmental excess charges in the nanowire structure is reduced by decreasing the excitation intensity. The experimentally obtained transverse exciton spin relaxation times can be reproduced by the unconventional motional narrowing theory. To determine the longitudinal exciton spin relaxation time, we performed full polarization measurement and Mueller Calculus. The longitudinal exciton spin relaxation times evaluated by the random polarization components are shorter than the recombination time and elongates with decreasing the emission energy. We believe that obtained results can contribute to the understanding of effective exciton spin relaxation mechanism in nanostructure.

This work was supported in part by the Grant-in-Aid for Young Scientists (B), No. 20760002 and GCOE-GSIST, Hokkaido University. M.K., N.A., and V.Z. acknowledges funding from NWO. H.S acknowledge Prof. S. Muto, Prof. S. Adachi, and Dr. R. Kaji for fruitful discussion.


\begin{thebibliography}{00}
\bibitem{Add3} In case of excitation condition (II), $X^0$ PL peak become the dominant with increasing excitation power of cw-TIS laser because the overflood carriers from InAsP NW-QD to InP NW also conpensate the internal fields.
\end{thebibliography}
\end{document}